\documentstyle[12pt]{article}

\font\twlgot =eufm10 scaled \magstep1
\font\egtgot =eufm8
\font\sevgot =eufm7

\font\twlmsb =msbm10 scaled \magstep1
\font\egtmsb =msbm8
\font\sevmsb =msbm7

\newfam\gotfam
\def\pgot{\fam\gotfam\twlgot}
\textfont\gotfam\twlgot
\scriptfont\gotfam\egtgot
\scriptscriptfont\gotfam\sevgot
\def\got{\protect\pgot}

\newfam\msbfam
\textfont\msbfam\twlmsb
\scriptfont\msbfam\egtmsb
\scriptscriptfont\msbfam\sevmsb
\def\Bbb{\protect\pBbb}
\def\pBbb{\relax\ifmmode\expandafter\Bb\else\typeout{You cann't use
Bbb in text mode}\fi}
\def\Bb #1{{\fam\msbfam\relax#1}}

\def\thebibliography#1{\bigskip\section*{\large
\bf References\\}\list
  {[\arabic{enumi}]}{\settowidth\labelwidth{#1}\leftmargin\labelwidth
    \advance\leftmargin\labelsep
    \usecounter{enumi}}
    \def\newblock{\hskip .11em plus .33em minus .07em}
    \sloppy\clubpenalty4000\widowpenalty4000
    \sfcode`\.=1000\relax}

\def\op#1{\mathop{{\it\fam0} #1}\limits}

\newcommand{\beq}{\begin{equation}}
\newcommand{\eeq}{\end{equation}}
\newcommand{\ben}{\begin{eqnarray}}
\newcommand{\een}{\end{eqnarray}}
\newcommand{\be}{\begin{eqnarray*}}
\newcommand{\ee}{\end{eqnarray*}}
\newcommand{\bea}{\begin{eqalph}}
\newcommand{\eea}{\end{eqalph}}

\newcommand{\gT}{{\got T}}

\newcommand{\cT}{{\cal T}}

\newcommand{\cL}{{\cal L}}

\newcommand{\cE}{{\cal E}}
\newcommand{\cH}{{\cal H}}

\newcommand{\bL}{{\bf L}}
\newcommand{\bH}{{\bf H}}

\newcommand{\dl}{\delta}

\newcommand{\f}{\phi}
\newcommand{\vf}{\varphi}

\newcommand{\Om}{\Omega}

\newcommand{\g}{\gamma}
\newcommand{\G}{\Gamma}

\newcommand{\th}{\theta}

\newcommand{\si}{\sigma}

\newcommand{\w}{\wedge}

\newcommand{\wt}{\widetilde}
\newcommand{\wh}{\widehat}

\newcommand{\dr}{\partial}

\newcommand{\ar}{\op\longrightarrow}

\newcommand{\ap}{\approx}

\newenvironment{eqalph}{\stepcounter{equation}
\setcounter{equationa}{\value{equation}}
\setcounter{equation}{0}

\begin{eqnarray}}{\end{eqnarray}\setcounter{equation}{\value{equationa}}}

\newcommand{\mar}[1]{}

\begin{document}
\hbox{}

{\parindent=0pt

{\large \bf Noether conservation laws in classical mechanics}
\bigskip 

{\bf G. Sardanashvily}

\medskip

\begin{small}

Department of Theoretical Physics, Moscow State University, 117234
Moscow, Russia

E-mail: sard@grav.phys.msu.su

URL: http://webcenter.ru/$\sim$sardan/
\bigskip

{\bf Abstract.}
In Lagrangian mechanics, Noether conservation laws including the
energy one are obtained similarly to those in field theory.
In Hamiltonian 
mechanics, Noether conservation laws 
are issued from the invariance of the Poincar\'e--Cartan integral invariant
under one-parameter groups of diffeomorphisms of a configuration space.
Lagrangian and Hamiltonian conservation laws need not be equivalent.
\end{small}
}

\bigskip
\bigskip

Classical non-relativistic mechanics can be formulated as a particular
field theory whose configuration space is a fibre bundle $Q\to \Bbb R$ 
over the time axis $\Bbb
R$ \cite{book98,book00,jmp00,sard98}. This configuration space is
equipped with bundle coordinates 
$(t,q^i)$ where
$t$ is the Cartesian coordinate on $\Bbb R$ possessing
transition functions $t'=t+$const., i.e., time reparametrizations
are not considered. Of course, a fibre bundle $Q\to\Bbb
R$ is trivial. However, it 
need not admit a preferable trivialization and, moreover, its
different trivializations correspond to different non-relativistic
reference frames. Therefore, we do not fix a trivialization of
$Q\to\Bbb R$ fixed, i.e., bundle coordinates $q^i$ possess arbitrary 
time-dependent transition functions $q^i\to q'^i(t,q^j)$.

We obtain Noether conservation laws in Lagrangian and Hamiltonian
non-relativistic mechanics similarly to those in field theory (see
\cite{book98} for a detailed exposition).

\section{Lagrangian conservation laws}

Following general formalism of dynamical systems on fibre bundles (see, e.g.,
\cite{book,epr}), 
the velocity phase space of non-relativistic mechanics on a
configuration space $Q$ is the first order
jet manifold $J^1Q$ of sections of $Q\to\Bbb R$. 
It is equipped with
the adapted coordinates $(t,q^i,q^i_t)$. There is the canonical
imbedding of $J^1Q \to TQ$ onto the subbundle of the tangent bundle
$TQ$ of $Q$ given by the condition $dt\rfloor v=1$, $v\in TQ$.
Due to this imbedding, any
connection $\G: Q\to J^1Y$ on a fibre bundle $Q\to \Bbb R$ is
represented by a vector field 
\mar{n0}\beq
\G=\dr_t +\G^i(t,q^j)\dr_i, \qquad dt\rfloor \G=1, \label{n0}
\eeq
on $Q$, and {\it vice versa}. Furthermore, one can associate to any 
connection $\G$ (\ref{n0}) an atlas of the fibre bundle $Q\to\Bbb R$ 
with time-independent transition functions such
that $\G=\dr_t$ with respect to the corresponding bundle coordinates.
In particular, there is one-to-one correspondence between the Ehresmann
connections represented by complete vector fields (\ref{n0}) 
and the trivializations of $Q\to\Bbb R$ \cite{book98,book00}.
From the physical viewpoint, any connection
$\G$ (\ref{n0}) defines a non-relativistic reference frame such that 
$q^i_t-\G^i$ are relative velocities with respect to this reference frame
\cite{book98,massa,sard98}.

A Lagrangian of non-relativistic mechanics is defined as a density
\mar{n1}\beq
L=\cL(t,q^j,q^j_t)dt \label{n1}
\eeq
on the velocity phase space $J^1Q$. 
Its variation $\dl L$ is the second order Euler--Lagrange operator
\mar{305}\beq
\cE_L=(\dr_i-d_t\dr^t_i)\cL \th^i\w dt, \label{305}
\eeq
where $d_t=\dr_t+q^i_t\dr_i +q^i_{tt}\dr^t_i$ is the total derivative and
$\th^i=dq^i-q^i_tdt$ are contact forms.
Its kernel Ker$\,\cE_L\subset J^2Q$ defines the Lagrange equation
\mar{n7}\beq
\cE_i=(\dr_i-d_t\dr_i^t)\cL=0. \label{n7}
\eeq
Here, $J^2Q$ is the second order jet manifold of sections of $Q\to\Bbb
R$ equipped with the adapted coordinates $(t,q^i,q^i_t,q^i_{tt})$. 

There are different Lagrangians whose variations provide the
same Euler--Lagrange operator. They make up an affine space
modelled over the vector space of variationally trivial Lagrangians
$L_0$, i.e., $\dl L_0=0$. One can show that a first order
Lagrangian $L_0$ is variationally trivial iff 
\mar{mos11}\beq
L_0=h_0(\vf)=(\vf_t+q^i_t\vf_i)dt, \label{mos11}
\eeq
where
$\vf=\vf_t dt +\vf_idq^i$ is a closed one-form on $Q$
and
\be
h_0(dt)=dt, \qquad h_0(\th^i)=0
\ee
is the horizontal operator acting on semibasic exterior forms on
$J^1Q\to Q$. 

A Noether conservation law results
from the invariance of a Lagrangian $L$ under a local one-parameter group of
bundle automorphisms of a configuration bundle $Q\to \Bbb R$
\cite{cari91,eche95,book98,mun01}. We agree to call them gauge
transformations by analogy with field theory. The
infinitesimal generator of a one-parameter gauge transformation group
is a projectable vector field 
\mar{n2}\beq
u=u^t\dr_t +u^i(t,q^j)\dr_i \label{n2}
\eeq
on $Q$, where $u^t=0,1$ because time reparametrizations are not considered.  
If $u^t=0$,
we have 
a vertical vector field $u=u^i\dr_i$ which takes its values into the
vertical cotangent bundle $VQ$ of $Q\to\Bbb R$. If $u^t=1$, a vector field 
$u$ (\ref{n2}) is a connection on the configuration 
bundle $Q\to\Bbb R$.
Connections $\G$ (\ref{n1}) on $Q\to\Bbb R$ make up an affine space
modelled over the vector space of vertical vector fields on $Q\to\Bbb
R$, i.e., the sum of a connection and a vertical vector field is a
connection, while the difference of two connections is a vertical
vector field on $Q\to\Bbb R$.   

The canonical jet prolongation of a vector field $u$ (\ref{n2}) onto
the velocity phase space  
$J^1Q$ reads
\mar{n3}\beq
J^1u=u^t\dr_t +u^i\dr_i + d_tu^i\dr_i^t. \label{n3}
\eeq
A Lagrangian $L$ (\ref{n1}) is invariant under a one-parameter group of
of gauge transformations generated by a vector field $u$
(\ref{n2}) iff its Lie derivative 
\mar{n23}\beq
\bL_{J^1u}L=J^1u\rfloor dL + d(u^t\cL)=(J^1u\rfloor d\cL)dt=
(u^t\dr_t +u^i\dr_i +d_tu^i\dr^t_i)\cL dt \label{n23}
\eeq
along $J^1u$ (\ref{n3})
vanishes.   
The first 
variational formula provides the canonical decomposition 
\mar{n4}\beq
\bL_{J^1u}L= (u^i-u^tq_t^i)\cE_i dt+ d_t (u\rfloor H_L) dt, \label{n4} 
\eeq
of the Lie
derivative (\ref{n23}), where 
\mar{n30}\beq
H_L=L+\dr^t_i\cL \th^i \label{n30}
\eeq
is the Poincar\'e--Cartan form. For instance, if $L=L_0$ is a
variationally trivial Lagrangian (\ref{mos11}), the first variational 
formula (\ref{n4}) gives the equality
\mar{v4}\beq
\bL_{J^1u}L_0=d_t(u\rfloor\vf)=d_t(u^t\vf_t +u^i\vf_i). \label{v4}
\eeq

If $\bL_{J^1u}L=0$, we have the Noether
conservation law
\mar{n6}\beq
0\ap-d_t\gT_u  \label{n6}
\eeq
of the symmetry function
\mar{n5}\beq
\gT_u=-u\rfloor H_L=(u^tq_t^i-u^i)\dr_i^t\cL- u^t\cL \label{n5}
\eeq
on the shell (\ref{n7}). 

Since $u^t=0,1$, there are the following two types of
Lagrangian conservation laws (\ref{n6}).

Let $u=v^i\dr_i$ be a vertical vector field on $Q\to \Bbb R$. If the
Lie derivative $\bL_{J^1 u}L$ vanishes, we obtain the Noether
conservation law
\mar{n20}\beq
d_t(v^i\dr^t_i\cL)\ap 0 \label{n20}
\eeq
of the momentum 
\mar{n21}\beq
\gT_v=-v^i\dr^t_i\cL \label{n21}
\eeq
along a vector field $u$. In particular, let $u(q)\neq 0$ at a point
$q\in Q$. Then, there exists an open neighbourhood $U$ of $q$ provided with
coordinates $q'^i$ such that $u=\dr/\dr q'^1$. A glance at the
expression (\ref{n23}) shows that the Lie derivative $\bL_{J^1u}L$ on
$U$ vanishes iff a Lagrangian $\cL(t,q'^j,q'^j_t)$ is independent of
the coordinate $q'^1$. 

Let $u=\G=\dr_t +\G^i\dr_i$ be a connection. The corresponding symmetry
function (\ref{n5}) is the energy function
\mar{n22}\beq
\gT_\G=(q^i_t-\G^i)\dr^t_i\cL-\cL \label{n22}
\eeq
\cite{eche95,book98,mun01}. If the Lie derivative $\bL_{J^1\G}L$
vanishes, we have the energy conservation law
\mar{n24}\beq
d_t((q^i_t-\G^i)\dr^t_i\cL-\cL)\ap 0. \label{n24}
\eeq
As was mentioned above, one can always choose bundle coordinates on 
$Q\to\Bbb R$ such that 
$\G^i=0$. A glance at the
expression (\ref{n23}) shows that the Lie derivative $\bL_{J^1\G}L$ 
vanishes iff a Lagrangian $L$ written with respect
to these coordinates is independent of time. Then, the energy
conservation law (\ref{n24}) takes the form
\be
d_t(q^i_t\dr^t_i\cL-\cL)\ap 0. 
\ee

We observe that there are different energy functions $\gT_\G$
(\ref{n22}) corresponding to different connections $\G$ on $Q\to\Bbb R$.
Moreover, if an energy function $\gT_\G$ is conserved and the momentum 
$\gT_v$ (\ref{n21}) is so, the energy function 
\be
\gT_{\G+v}=\gT_\G+\gT_v
\ee
is also conserved. Therefore, the problem is to select an energy
function describing a true physical energy. Simple examples show that
it may be an energy function $\gT_\G$ where a connection $\G$ takes its
values into the kernel of the Legendre map (\ref{n31}). The problem is
that, if a Lagrangian $L$ is degenerate, such a connection fails to be
unique. 

{\bf Example.} Let us consider a one-dimensional motion of a point particle
subject to friction. It is described by the dynamic equation
\be
q_{tt}=-kq_t, \qquad k>0.
\ee
This equation is equivalent to the Lagrange equation of the Lagrangian
\mar{n80}\beq
L=\frac12 \exp[kt]q^2_t dt. \label{n80}
\eeq
Let us consider the vector field
\be
\G=\dr_t -\frac{k}{2}q\dr_q.
\ee
Its jet prolongation (\ref{n3}) reads
\be
J^1\G=\dr_t -\frac{k}{2}q\dr_q -\frac{k}{2}q_t\dr_q^t.
\ee
It is readily observed that the Lie derivative of the Lagrangian (\ref{n80})
along $J^1\G$ vanishes. Then, the energy function
\be
\gT_\G=\frac12\exp[kt]q_t(q_t+kq)
\ee 
is conserved.

It however may happen that the Lie derivative of a Lagrangian does not vanish,
but a conservation law takes place as follows. 

(i) Every Lagrangian $L$ (\ref{n1}) yields the Legendre map
\mar{n31}\beq
\wh L: J^1Q\ar_Q V^*Q, \qquad p_i\circ \wh L=\dr_i^t\cL, \label{n31}
\eeq
of the velocity phase space $J^1Q$ to the vertical cotangent bundle
$V^*Q$ of $Q\to \Bbb R$ equipped with the holonomic coordinates
$(t,q^i,p_i)$. This bundle plays a role of the momentum phase space of
non-relativistic mechanics (see next Section). A Lagrangian $L$ is
called regular if the Legendre map (\ref{n31}) is of maximal rank,
i.e., a local diffeomorphism. Otherwise, it is said to be degenerate.
Let $L$ be a degenerate Lagrangian and $v=v^i\dr_i$ a vertical vector
field on $Q\to \Bbb R$ which belongs to the annihilator Ann$(\wh
L(J^1Q))\subset VQ$ of $\wh L(J^1Q)$, i.e., 
\be
v^i(t,q^j)\dr^t_i\cL(t,q^j,q^j_t)=0.
\ee
In this case, the first variational formula (\ref{n4}) takes the form
\be
\bL_{J^1v}L= u^i\cE_i dt,
\ee
i.e., the Lie derivative $\bL_{J^1v}L$ vanishes on the shell (\ref{n7}).
Let now $u$ be a vector field such that $\bL_{J^1u}L=0$, and let us
consider the sum $u+v$. The Lie derivative $\bL_{J^1(u+v)}L$ does not vanish,
but it vanishes on-shell. Applying to it the first variational formula 
(\ref{n7}), we recover the Noether conservation law (\ref{n6}).

(ii) The Euler-Lagrange operator $\cE_L$ (\ref{305})
is invariant under a one-parameter group
of gauge transformations generated by a vector field $u$
iff its Lie derivative  $\bL_{J^2u}\cE_L$ along the
jet prolongation $J^2u$ of $u$ onto the second order jet manifold $J^2Q$ 
vanishes. There is the relation
\mar{v3}\beq
\bL_{J^2u}\cE_L=\dl(\bL_{J^1u}L)=\cE_{\bL_{J^1u}L}, \label{v3}
\eeq 
i.e., the Lie derivative $\bL_{J^2u}\cE_L$ is the Euler--Lagrange
operator associated to the Lagrangian $\bL_{J^1u}L$ \cite{book}.
It follows that 
$\bL_{J^2u}\cE_L=0$ iff the Lie derivative $\bL_{J^1u}L$ is a
variationally trivial Lagrangian (\ref{mos11}), i.e.,
\mar{n46}\beq
\bL_{J^1u}L=h_0(\vf), \label{n46}
\eeq
where $\vf$ is a closed form on $Q$. 
Substituting this expression into the first variational formula
(\ref{n4}), we obtain the weak equality
\mar{v5}\beq
h_0(\vf)\ap d_t(u\rfloor H_L)\label{v5}
\eeq
on the shell (\ref{n7}). If $\vf=d\si$ is an exact form on $Q$, 
this equality is
brought into the weak conservation law
\mar{v6}\beq
0\ap d_t (u\rfloor H_L-\si) \label{v6}
\eeq
due to the relation $d_t\circ h_0=h_0\circ d$.
Let $L'=L+L_0$ be another Lagrangian whose variation is the
Euler--Lagrange operator $\cE_L$. Due to the equality (\ref{v4}), we
come to the same conservation law (\ref{v6}).
Note that, if $\bL_{J^1u}L=h_0(\vf)$,
the Lie derivative of the Poincar\'e--Cartan form $H_L$ (\ref{n30}) is
\mar{n40}\beq
\bL_{J^1u}H_L=J^1u\rfloor dH_L + d(J^1u\rfloor H_L)=\vf, \label{n40}
\eeq
and {\it vice versa}. This fact
follows at once from the equality
\mar{n33}\beq
\bL_{J^1u}H_L= \bL_{J^1u}L + \dr^t_i(J^1u\rfloor d\cL)\th^i.
\label{n33}
\eeq

{\bf Example.} Let us consider a one-dimensional free motion of a point
particle described by the Lagrangian
\mar{n82}\beq
L=\frac12q^2_tdt. \label{n82}
\eeq
The vector field $u=vt\dr_q$, $v=$const., is the infinitesimal generator of a
one-parameter group of the Galilei transformations. Its jet
prolongation (\ref{n3}) reads
\be
J^1u=vt\dr_q +v\dr_q^t.
\ee
The Lie derivative of the free motion Lagrangian (\ref{n82}) along
$J^1u$ is 
\be
\bL_{J^1u}L=vq_t=d_t(vq).
\ee
Then, the equality (\ref{v6}) shows that $(q_tt-q)$ is a constant of motion.

\section{Hamiltonian conservation laws}

As was mentioned above, the momentum phase space of non-relativistic
mechanics on a
configuration space $Q$ is the vertical
cotangent bundle $V^*Q$ of $Q\to\Bbb R$ equipped with holonomic coordinates 
$(t,q^i,p_i)$. 
The cotangent
bundle $T^*Q$ of $Q\to\Bbb R$ coordinated by $(t,q^i,p,p_i)$ plays a role
of the homogeneous momentum phase space. 
It is provided with the canonical Liouville form $\Xi=pdt +p_idq^i$
and the canonical symplectic form $\Om=d\Xi$.

There are three equivalent description of the dynamics of
non-relativistic Hamiltonian mechanics.

(i) There is the trivial affine
bundle 
\mar{n9}\beq
\zeta: T^*Q \to V^*Q. \label{n9}
\eeq
Its section 
\mar{n10}\beq
h: V^*Q\to T^*Q, \qquad p\circ h=-\cH(t,q^i,p_i) \label{n10}
\eeq
is a Hamiltonian of (time-dependent) non-relativistic mechanics.  
The pull-back 
$h^*\Xi$ of $\Xi$ onto $V^*Q$ by means of a section $h$ (\ref{n10}) is the
well-known Poincar\'e--Cartan integral invariant
\mar{n11}\beq
H=p_idq^i-\cH dt. \label{n11}
\eeq
We agree to call it a Hamiltonian form. 
There exists a unique vector field $\g_H$ on $V^*Q$ such that
\mar{n12}\ben
&& dt\rfloor \g_H=1, \qquad \g_H\rfloor dH=0, \nonumber\\
&& \g_H=\dr_t+\dr^i\cH\dr_i-\dr_i\cH\dr^i. \label{n12}
\een
It defines the first order Hamilton equation
\mar{n13}\beq
d_t q^i=\dr^i\cH, \qquad d_tp_i=-\dr_i\cH \label{n13}
\eeq
on $V^*Q$, where $d_t=\dr_t +q^i_t\dr_i + p_{ti}\dr^i$ is the
total derivative written with respect to the adapted
coordinates $(t,q^i,p_i,q_t^i,p_{ti})$ on 
the jet manifold $J^1V^*Q$ of the fibre bundle $V^*Q\to\Bbb R$.

(ii) Let us take the pull-back of a Hamiltonian
form $H$ (\ref{n11}) onto $J^1V^*Q$, and let us 
consider the Lagrangian
\mar{n41}\beq
L_H=h_0(H)=(p_iq^i_t-\cH(t,q^j,p_j))dt \label{n41}
\eeq
on $J^1V^*Q$. It is readily observed that the Lagrange equation of $L_H$
is exactly the Hamilton equation (\ref{n13}).

(iii) Let us consider the pull-back $\zeta^*H$ of the Hamiltonian form
$H$ (\ref{n11}) onto $T^*Q$. Then
\mar{n70}\beq
\bH=\dr_t\rfloor (\Xi-\zeta^*H)=p+\cH
\eeq
is a function on $T^*X$. Let us regard it as a Hamiltonian of
an autonomous Hamiltonian system on the symplectic manifold $(T^*Q,\Om)$
provided with the corresponding Poisson bracket
\be
\{f,f'\}_T=\dr^pf\dr_tf' +\dr^if\dr_if'-
\dr_tf\dr^pf' -\dr_if\dr^if', \qquad \dr^p=\dr/\dr p.
\ee
Then, the relation
\mar{n72}\beq
\zeta^*(\bL_{\g_H}f)=\{\bH,\zeta^*f\}_T \label{n72}
\eeq
holds for any smooth real function $f\in C^\infty(V^*Q)$. For instance,
$f$ is a first integral of motion iff the bracket $\{\bH,\zeta^*f\}_T$
vanishes. 

Turn now to Noether conservation laws. 
Let $u$ be a projectable vector field (\ref{n2}) on 
$Q\to\Bbb R$ treated as the generator of a one-parameter group of gauge
transformations. Its canonical lift onto $V^*Q$ reads
\mar{n8}\beq
\wt u=u^t\dr_t +u^i\dr_i-p_j\dr_i u^j\dr^i. \label{n8}
\eeq
The Lie derivative of the Hamiltonian form $H$ (\ref{n11}) along $\wt
u$ (\ref{n8}) reads
\mar{n42}\beq
\bL_{\wt u}H=\wt u\rfloor dH+ d(\wt u\rfloor H)=(\dr_t(p_iu^i-u^t\cH) -
u^i\dr_i\cH +\dr_i u^j p_j\dr^i\cH)dt.
\label{n42}
\eeq
Let $J^1\wt u$ be the jet prolongation of $\wt u$ onto $J^1V^*Q$. A
simple computation shows that the pull-back of the Lie derivative
(\ref{n42}) onto $J^1V^*Q$ obeys the relation
\mar{n43}\beq
\bL_{\wt u}H = \bL_{J^1\wt u}L_H. \label{n43}
\eeq
In particular, the Hamiltonian form $H$ (\ref{n11}) is
invariant under a one-parameter group of gauge transformations iff the
Lagrangian $L_H$ (\ref{n41}) is so.

Regarding the Hamilton equation (\ref{n13}) as the Lagrange equation
of the Lagrangian $L_H$ (\ref{n41}), we have the condition (\ref{n46})
of its
invariance under a one-parameter group of gauge transformations
generated by a vector field $u$ (\ref{n2}), i.e., 
\be
\bL_{J^1\wt u}L_H=\bL_{\wt u}H=h_0(\f), 
\ee
where $\f$ is a closed form on $V^*Q$. Moreover, since the left-hand side
of this relation is independent of jet coordinates $q^i_t$ and $p_{ti}$,
this condition takes the form
\mar{n45}\beq
\bL_{J^1\wt u}L_H=\bL_{\wt u}H=\dr_tf(t)dt, \label{n45}
\eeq
where $f$ is a function of time only. Then, applying the first
variational formula (\ref{n23}) to the Lagrangian $L_H$ (\ref{n41}),
one can obtain Noether conservation laws in Hamiltonian mechanics
\cite{book98,jmp00,sard98}. 
This first variational formula reads
\mar{n47}\beq
\bL_{J^1\wt u}L_H=-[(u^i-u^tq^i_t)(p_{ti}+\dr_i\cH)+ (p_j\dr_i u^j +
u^tp_{ti})(q^i_t-\dr^i\cH) + d_t(u^t\cH-u^ip_i)]dt. \label{n47}
\eeq
If the Lie derivative $\bL_{J^1\wt u}L_H$ obeys the relation
(\ref{n45}), we obtain the conservation law
\mar{n48}\beq
0\ap - d_t(u^t\cH-u^ip_i+f) \label{n48}
\eeq
on the shell (\ref{n13}). If the Lie derivative (\ref{n47}) vanishes,
this conservation law takes the form
\mar{n49}\beq
0\ap - d_t(u^t\cH-u^ip_i). \label{n49}
\eeq
We agree to call
\mar{n50}\beq
\cT_u=u^t\cH-u^ip_i \label{n50}
\eeq
a symmetry function. In next Section, we will relate it to a symmetry
function in Lagrangian mechanics.

Equivalently, the conservation law (\ref{n49}) results from the
equality
\be
\bL_{\wt u}H= -\g_H\rfloor d\cT_u
\ee
and from the fact that $d_tf=\g_H\rfloor df$ on-shell for any function $f$
on $V^*Q$. Then, it follows from the relation (\ref{n72}) that a
conserved symmetry function $\cT_u$ is a first integral.

In particular, let $u=v^i\dr_i$ be a vertical vector field on $Q\to
\Bbb R$. If the 
Lie derivative $\bL_{J^1\wt u}L_H$ vanishes, we obtain the Noether
conservation law
\mar{n51}\beq
d_t(v^ip_i)\ap 0 \label{n51}
\eeq
of the momentum 
\mar{n52}\beq
\cT_v=-v^ip_i \label{n52}
\eeq
along a vector field $u$. Let $u(q)\neq 0$ at a point
$q\in Q$. As was mentioned above, there exists an open neighbourhood
$U$ of $q$ provided with 
coordinates $q'^i$ such that $u=\dr/\dr q'^1$. A glance at the
expression (\ref{n42}) shows that the Lie derivative $\bL_{J^1\wt u}L_H$ on
$U$ vanishes iff $\cH(t,q'^j,p'_j)$ is independent of
the coordinate $q'^1$. 

Let $u=\G=\dr_t +\G^i\dr_i$ be a connection. The corresponding symmetry
function (\ref{n50}) is the energy function
\mar{n53}\beq
\cT_\G=\cH_\G=\cH-p_i\G^i. \label{n53}
\eeq
If the Lie derivative $\bL_{J^1\wt\G}L_H$
vanishes, we have the energy conservation law
\mar{n54}\beq
d_t\cH_\G\ap 0. \label{n54}
\eeq
As was mentione above, one can always choose bundle coordinates on 
$Q\to\Bbb R$ such that 
$\G_i=0$ and $\cH=\cH_\G$. A glance at the
expression (\ref{n42}) shows that the Lie derivative $\bL_{J^1\wt\G}L_H$ 
vanishes iff the energy $\cH_\G$ written with respect
to these coordinates is independent of time.

\section{Relations between Lagrangian and Hamiltonian conservation laws}

Lagrangian and Hamiltonian formulations of non-relativistic mechanics
fail to be equivalent. The relationship between Lagrangian and
Hamiltonian mechanics \cite{book98,jmp00} is a particular case of that
between Lagrangian and Hamiltonian formulations of field theory
\cite{book,jpa}. 

As was mentioned above, every Lagrangian $L$ (\ref{n1}) on the velocity
phase space $J^1Q$ yields the Legendre map $\wh L$ (\ref{n31}). 
Conversely, any Hamiltonian form $H$ (\ref{n11}) on
the momentum phase space $V^*Q$ yields the momentum map
\mar{n60}\beq
\wh H: V^*Q\ar_Q J^1Q, \qquad q^i_t\circ\wh H=\dr^i\cH. \label{n60}
\eeq

Given a Lagrangian $L$ on $J^1Q$, 
a Hamiltonian form $H$ on $V^*Q$ is said to be
associated with $L$ if $H$ satisfies the relations
\mar{2.30,1}\ben
&&\wh L\circ\wh H\circ \wh L=\wh L,\label{2.30} \\
&& \cH=p_i\dr^i\cH -\wh H^*\cL, \label{2.31}
\een
where $\wh H^*\cL=\cL(t,q^j,\dr^j\cH)$ is the pull-back of $\cL$ onto $V^*Q$.
An associated Hamiltonian form need not
exists or it is not necessarily 
unique. Here, we will restrict our consideration to the cases of 
hyperregular Lagrangians (see \cite{book98} for the case of semiregular
Lagrangians). 

A Lagrangian $L$ is said to be hyperregular if the Legendre map $\wh L$
is a bundle isomorphism of $J^1Q\to Q$ onto $V^*Q\to Q$. 
In this case, there exists a unique associated
Hamiltonian form $H$ such that: (i) the momentum map $\wh H=\wh L^{-1}$ is the 
inverse bundle isomorophism, (ii) the Hamiltonian form $H$ is 
the pull-back $H=\wh H^*H_L$ 
of the Poincar\'e--Cartan form $H_L$, and (iii) 
the Poincar\'e--Cartan form $H_L$ is the pull-back
$H_L=\wh L^* H$ of $H$.
The property (i) takes the coordinate form
\mar{n86}\beq
p_i=\dr_i^t\cL(t,q^j,\dr^j\cH(t,q^k,p_k)), \qquad
q^i_t=\dr^i\cH(t,q^j,\dr_j^t\cL(t,q^k,q^k_t)). \label{n86}
\eeq
The property (ii) recovers the relation (\ref{2.31}). The property (iii)
leads to the coordinate equality
\mar{n87}\beq
q^i_t\dr^t_i\cL(t,q^j,q^j_t)-\cL(t,q^j,q^j_t)=\cH(t,q^j,
\dr^t_j\cL(t,q^k,q^k_t)). \label{n87}
\eeq
Furthermore, one can show that, if $\wt s(t)=(q^i(t), p_i(t))$ is a solution of
the Hamilton equation (\ref{n31}) of $H$, then $s(t)=(q^i(t))$ is a solution
of the Lagrange equation (\ref{n7}) of $L$. Conversely, if 
$s(t)$ is a solution
of the Lagrange equation (\ref{n7}) of $L$, then 
\mar{n89}\beq
\wt s(t)=(\wh L\circ s)(t)=(q^i(t),
p_i(t)=\dr_i^t\cL(t,q^j(t), \dr_tq^j(t))) \label{n89}
\eeq
 is a solution of the
Hamilton equation (\ref{n31}) of $H$. Thus, it seems that Lagrangian
and Hamiltonian formalisms in the case of hyperregular Lagrangians are
completely equivalent. However, it
appears that the relation between Lagrangian and Hamiltonian
conservation laws is more intricate. 

Let $u$ be a projectable vector field (\ref{n2}) on $Q$, and let $J^1u$
(\ref{n3}) and $\wt u$ (\ref{n8}) be its prolongations onto $J^1Q$ and
$V^*Q$, respectively. The key point is that, though the Hamiltonian
form $H$ and the Poincar\'e--Cartan form $H_L$ are the pull-back of
each other, their Lie derivatives $\bL_{\wt u}H$ and $\bL_{J^1u}H_L$ 
fail to be so because the vector fields $J^1u$
and $\wt u$ are not transformed into each other by morphisms $\wh L$
and $\wh H$. 
At the same time, using the formulas (\ref{2.31}) -- (\ref{n87}), one can
obtain the relations 
\be
\wh H^*\gT_u=\cT_u, \qquad \wh L\cT_u=\gT_u
\ee
between the symmetry functions $\gT_u$ (\ref{n5}) and $\cT_u$
(\ref{n50}). It follows that the symmetry function $\gT_u$ is constant
on a solution $s$ of the Lagrange equation, i.e., $\dr_ts^*\gT_u=0$ iff
the symmetry function $\cT_u$ is constant on the corresponding
solution $\wt s$ (\ref{n89}) of the Hamilton equation, and {\it vice versa}.

In general case, there is no one-to-one correspondence between
solutions of the Lagrange and Hamilton equations. For instance, it may
happen that different solutions $s$ and $s'$ of the Lagrange equation
of a Lagrangian $L$ are associated to solutions $\wt s$ and $\wt s'$
of the Hamilton equations 
of different Hamiltonian forms $H$ and $H'$ associated with $L$ 
such that we have different symmetry functions constant on $\wt s$ and
$\wt s'$ \cite{book98}.

\end{document}